# High-energy kink in high-temperature superconductors

T. Valla[1*], T. E. Kidd[1,2], Z.-H. Pan[3], A. V. Fedorov[3], W.-G. Yin[1], G.D. Gu[1] and P.D. Johnson[1]

[1] *Condensed Matter Physics and Materials Science Department, Brookhaven National Laboratory, Upton, NY 11973, USA.*

[2] *Physics Department, University of Northern Iowa, CedarFalls, IA 50614-0150, USA*

[3] *Advanced Light Source, Lawrence Berkeley National Laboratory, Berkeley, CA 94720, USA.*

* e-mail:valla@bnl.gov

In conventional metals, electron-phonon coupling, or the phonon-mediated interaction between electrons, has long been known to be the pairing interaction responsible for the superconductivity. The strength of this interaction essentially determines the superconducting transition temperature $T_C$. One manifestation of electron-phonon coupling is a mass renormalization of the electronic dispersion at the energy scale associated with the phonons. This renormalization is directly observable in photoemission experiments.[1] In contrast, there remains little consensus on the pairing mechanism in cuprate high temperature superconductors. The recent observation of similar renormalization effects in cuprates has raised the hope that the mechanism of high temperature superconductivity may finally be resolved.[2,3,4,5,6,7,8,9] The focus has been on the low energy renormalization and associated "kink" in the dispersion at around 50 meV. However at that energy scale, there are multiple candidates including phonon branches, structure in the spin-fluctuation spectrum, and the superconducting gap itself, making the unique

**identification of the excitation responsible for the "kink" difficult. Here we show that the low-energy renormalization at ~50 meV is only a small component of the total renormalization, the majority of which occurs at an order of magnitude higher energy (~350 meV). This high energy "kink" poses a new challenge for the physics of the cuprates. Its role in superconductivity and relation to the low-energy "kink" remains to be determined.**

Fig. 1 shows wide energy range angle-resolved photoemission (ARPES) spectra for two different families of cuprate high-temperature superconductors (HTSC) at different doping levels: from optimally doped ($T_C$ = 91K) and highly underdoped ($T_C$ = 5K) $Bi_2Sr_2CaCu_2O_{8+\delta}$ (BSCCO) and from $La_{2-x}Ba_xCuO_4$ (LBCO) at x=0.095 ($T_C$ = 32 K) and x=1/8 ($T_C$ = 2.5 K). Spectra were taken from several different lines in *k*-space at ~10 K, in the normal state for underdoped BSCCO and x=1/8 LBCO and in the superconducting state for optimally doped BSCCO and LBCO at x=0.095. In the latter cases, the familiar low energy kink can easily be seen at approximately 50 meV. However, another, far more pronounced kink can be seen in all the samples at much higher binding energies. Measurements of the intensity as a function of momentum or momentum distribution curves (MDC) show relatively well-defined peaks down to 0.8 eV that can be fitted with Lorentzian line-shapes. The MDC derived dispersions clearly show "kinks" at high energies in the range around 300-400 meV, while at even higher energies (~0.7 eV) they tend to recover the bare tight-binding (TB) dispersions.[10] In the energy range between these two limits, the MDC peaks show very steep, and in the BSCCO case, virtually "vertical dispersion".

There are two recent reports showing similar results in the BSCCO family of cuprates.[11,12] The authors of one paper[11] suggest that the observed effect is not a renormalization, but rather a shift of the spectral weight that produces an apparent "vertical dispersion" at the new zone boundary due to some ordering. In BSCCO samples it appears that the "vertical dispersion" or a "waterfall" occurs at momenta commensurate with ($\pi/4,\pi/4$) suggesting that some form of antiferromagnetism (AF) and corresponding four-folding of the Brillouin zone might be involved in producing the observed effects. Our results confirm that in BSCCO samples the "vertical dispersion" occurs near $(\pi/4,\pi/4)$. The "vertical dispersion" also appears in the LBCO system in the majority of momentum space, except near the nodal line. In the latter region, the rate of dispersion increases at high energies, but never becomes "vertical". However, we note that the wave-vector associated with the "vertical dispersion" is not common to all the cuprates – rather, it depends on doping and on details of the relevant "band structure".

This observation is illustrated in Fig. 2 where we show in (a) and (b) the spectral intensity at the Fermi level for two different doping levels of LBCO system. The x=0.165 sample is superconducting with $T_C$=24K. In (c) and (d) we show the corresponding contours at $\omega$=−0.4 eV. The latter contours represent the momentum positions corresponding to the "vertical dispersion". The contours stay nearly fixed in the energy range between 300 - 450 meV (except in the vicinity of the nodal line) but differ from one doping level to another and are all far from the $(\pi/4,\pi/4)$ contour, characteristics of BSCCO samples. In figure 2(e) we show the momentum $k_0$ characterizing the contours for the measured LBCO samples and compare it with corresponding $k_0$ magnitudes from the previously reported BSCCO[11] samples and from an undoped parent compound $Ca_2CuO_2Cl_2$ (CCOC) where the high energy kink has been observed at even slightly higher energy ~450 meV measured from the top of valence band.[13] This latter system is an antiferromagnetic correlated insulator with well defined magnetic excitations or magnons extending up to several hundred meV.

The shift towards $(\pi/2,\pi/2)$, or much larger $k_0$-vectors in the 214 (LBCO and $La_{2-x}Sr_xCuO_4$ (LSCO)) samples relative to BSCCO seems to be a simple consequence of the different bare dispersions and Fermi surface shapes for these two families of cuprates. Due to a smaller next nearest neighbour hopping $t'$ in the 214 systems the constant energy contours at low energies are less curved and tend to be closer to the $(\pi/2,\pi/2)$ point than in BSCCO family. Therefore, the "vertical dispersion" is pushed closer to $(\pi/2,\pi/2)$ in the 214 systems. This suggests that the observed high-energy kink is not an "ordering effect" but that it rather represents a true renormalization and opens the possibility of extracting the complete self-energy.

The real and imaginary components of the self-energy are derived in the usual manner from the widths, $\Delta k$, and peak positions, $k_m$, of the MDC peaks using the expressions

$$k_m = k_F + [\omega - \text{Re}\Sigma(\omega)]/v_0 \quad \text{and} \quad \Delta k = 2\,\text{Im}\Sigma(\omega)/v_0 \qquad (1)$$

The MDCs are Lorentzians if the bare dispersion is linear, $\varepsilon_k = v_0(k - k_F)$, and $\Sigma$ does not vary strongly with momentum, an approximation that work well in the cuprates. Here $\text{Re}\Sigma(\omega)$ and $\text{Im}\Sigma(\omega)$ represent the real and imaginary components of the self energy at a binding energy $\omega$ and $v_0$ represents the bare velocity. The results are shown in Fig. 3. Panels a) and b) show the wide-range $\text{Re}\Sigma(\omega)$ for different doping levels of BSCCO and LBCO, respectively. For the non-interacting dispersions indicated in Fig. 1, we have used linear approximations based on the TB derived Fermi velocities for each probed line.[10] The resulting self-energies shown in Fig. 3 are almost one order of magnitude larger than those previously reported in studies focused on the low-energy renormalization and have dominant structures at ~340 meV. We note that the measured self-energies show several interesting characteristics: the high energy kink in both systems strengthens away from the nodal line, in accordance with ref. [12]. However, in contrast to BSCCO where only the amplitude varies, while the overall shape of $\text{Re}\Sigma$ remains nearly $k$-independent,[12] in the LBCO system the peak position and the shape of $\text{Re}\Sigma$ depend on $k$. Further, for a given momentum line, the high energy renormalization is stronger in BSCCO than in LBCO and in both systems only marginally dependent on doping.

The presence of a kink at these high energies immediately raises the question: what type of excitation spectrum is required to produce such a renormalization effect? We note that the phonon distribution in these materials is limited to ~80 meV. Furthermore the superconducting gap and the so called "pseudogap" are doping dependent and usually always less than 50 meV in $Bi_2Sr_2CaCu_2O_{8+\delta}$, and even smaller in LBCO.[14] Therefore, both phonons and (pseudo)gaps can clearly be ruled out as the origin of the large renormalization observed at high energies. The only excitations that are available that extend to these energies are spin fluctuations, or magnons, in the undoped material.[15,16,17,18]

In Fig. 4(a) we present a model distribution of "bosonic excitations" that would allow the reproduction of the self-energy characteristics presented in Fig. 3 for the optimally doped BSCCO in the superconducting state. Essentially, three features are required: a relatively narrow peak at low energy ~50-70meV, a rather broad peak centered at ~340 meV, and a continuum in the region in between. We note in passing that the narrow peak at low energies accounts for approximately 10% of the total spectrum. The electronic self-energies derived from such an excitation spectrum, using the standard momentum-independent Eliashberg formulae, are shown on the same scale with experimentally derived self-energies in figs 4(b) and 4(c). Note that the Im$\Sigma$ shown in fig. 4(c) saturates at the higher energies as Re$\Sigma$ passes through its peak. This is in a sharp contrast with previously reported behaviour, where Re$\Sigma$ peaked at ~50-70meV while Im$\Sigma$ had a linear dependence up to the highest measured energies, clearly violating the Kramers-Kronig causality relations. We note that the self energies shown in Fig. 4 approximately satisfy the Kramers-Kronig relations, suggesting that the TB bands represent good approximations to the bare dispersions and that the self-energies derived here have physical meaning.

Recently, neutron scattering experiments have been performed on $La_{2-x}Ba_xO_4$,[16] and on underdoped YBCO,[17,18] up to high energies (~300 meV), showing that the spin excitations are remarkably similar among different families of cuprate superconductors, consisting of a commensurate ($\pi,\pi$) scattering at some finite energy $\Omega_{res}$, and scattering branches dispersing downwards and upwards out of this ($\pi,\pi$) mode with increasing incommensurability. These experiments provide a unique opportunity to directly compare spin fluctuation spectra with the electronic self-energies within the full energy range relevant for spin dynamics for the first time. In Fig. 4 (e) and (f) we compare the measured self-energies for LBCO at x=1/8 with those modelled by using the $La_{2-x}Ba_xO_4$ susceptibilities from ref. [16]. The latter is shown in figure 4(d) as a solid line. We also show in figure 4(d) a modification to the excitation spectrum that better reproduces the measured self energies at high energies (dashed lines). We note that, similar to BSCCO and CCOC, additional weight is always needed at energies that go somewhat beyond a typical "single-magnon" (spin-fluctuation) continuum ~2-3J, suggesting that scattering on the two-magnon continuum might well make a contribution.

As for the BSCCO self-energies near optimal doping we note that the low energy (≤50meV) spin susceptibility in these systems is markedly temperature dependent. In the superconducting state, the spin spectrum is gapped and relatively well-defined modes and a strong commensurate "resonance" exist, while in the normal state, or above the pseudogap temperature, in the underdoped systems,[19] the spin gap closes, and the excitations get overdamped and lose identity. Due to these changes, the quasiparticle self-energies at low energies are also temperature dependent as shown in Fig. 3a for optimally doped BSCCO and in general agreement with previous experiments.[6,20,21]

The overall evidence shows that the high energy kink seen in photoemission most likely results from the high-energy spin excitations. Phonons do not exist at this energy scale, and magnetic interactions are very strong in this material. The relevant high-energy spin excitations are the short wavelength nearest-neighbour spin-flip transitions. Experimentally it has been shown that in the undoped material such high energy excitations are characterized by the q-vectors on the magnetic Brillouin zone boundary, i.e., the $(\pm\pi, 0)$ to $(0, \pm \pi)$ lines.[15] The coupling of quasiparticles to spin excitations with these energies/momenta is shown[22,23,24] to be proportional to $\cos(k_x - q_x) + \cos(k_y - q_y)$. The nodal line will therefore couple preferentially to $q=(\pi/2,\pi/2)$ and $q=(-\pi/2, -\pi/2)$, while on moving away from the nodal direction, coupling to $q= (\pm\pi, 0)$ and $q= (0, \pm \pi)$ excitations becomes predominant. However, as the hole at ~300 meV must end up at the Fermi surface, the scattering into the antinodal regions [$k\approx(\pm\pi,0)$ and $k\approx(0,\pm\pi)$] will always be dominant and will become stronger on moving away from the nodal line as is indeed observed here in LBCO (Fig. 3b) and in ref. [12] in BSCCO.

The data also reinforce the possibility that the low energy kink in the superconducting state reflects the formation of the resonant mode within the spin fluctuation spectrum. If the high energy spin-fluctuations, related to the nearest-neighbour spin-flip transitions, are able to produce the high-energy renormalization, then it is evident that the small structures at the low-energy side of the spin-fluctuation spectrum, that develop upon doping and have a pronounced temperature dependence, can produce the low-energy kink. Then, in agreement with previous studies,[6,20,21] only the low energy part of the single-particle self-energy shows a significant temperature and doping dependence (Fig. 3a).

## *Methods*

The experiments were carried out on a Scienta SES-2002 electron spectrometer at beam line U13UB of the National Synchrotron Light Source and on a Scienta SES-100 spectrometer at beam line 12.0.1 of the Advanced Light Source. The combined instrumental energy resolution was ~ 20 meV (~35meV) at 21 (110) eV photon energy used at U13UB (12.0.1), and the momentum resolution was ±0.004 (±0.015) Å$^{-1}$, correspondingly. Samples, grown by travelling solvent floating zone method, were mounted on a liquid He cryostat and cleaved *in-situ* in the UHV chamber with a base pressure $2 \times 10^{-9}$ Pa. The temperature was measured using a calibrated silicon sensor mounted near the sample. The photoemission spectra were analysed using momentum distribution curves (MDC).

## *Acknowledgements*

We acknowledge useful discussions with Andrey Chubukov, Maurice Rice and John Tranquada. The research work described in this paper was supported by the Office of Science, US Department of Energy.

Correspondence and requests for materials should be addressed to T.V.

## *Competing financial interests*
The authors declare that they have no competing financial interests.

*Figures:*

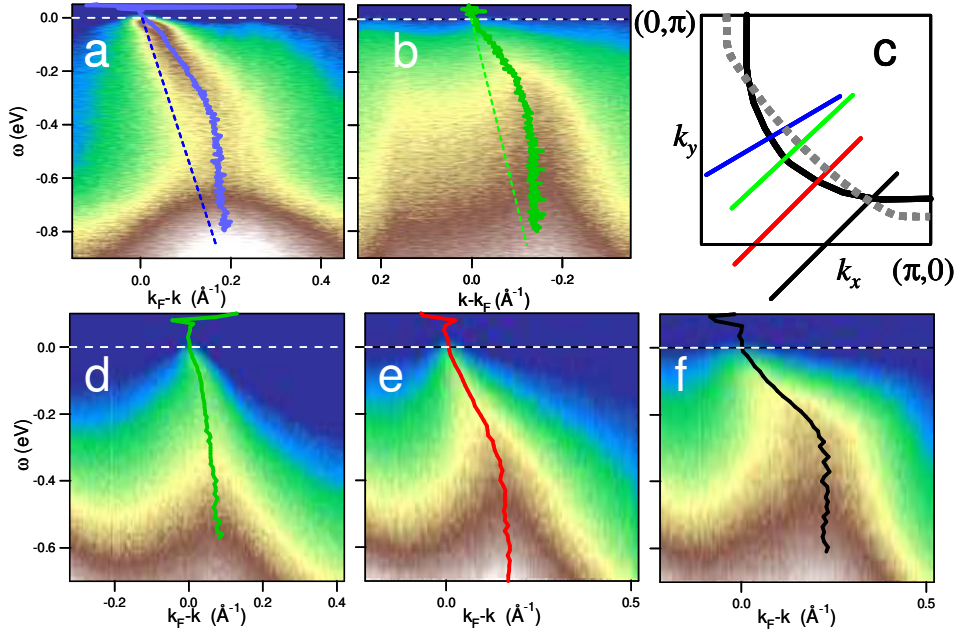

**Figure 1. Wide range photoemission spectra from different cuprates. a)** optimally doped BSCCO. **b)** very underdoped ($T_C$~5K) BSCCO. **c)** Brillouin zone with the Fermi surfaces for BSCCO (solid line) and LBCO (dashed line). Straight lines represent the momentum lines probed in the spectra with correspondingly colored dispersion. **d)** nodal LBCO spectrum for x=0.095. **e)** and **f)** LBCO spectra at x=0.125 for two different momentum lines as indicated in c). Dashed lines in the spectra represent bare tight-binding dispersions while solid lines represent MDC derived dispersions.

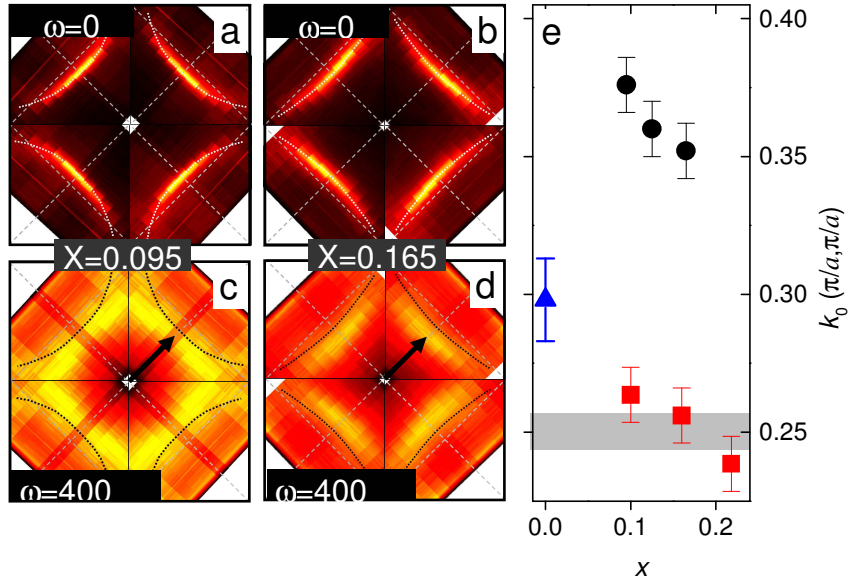

**Figure 2. Characteristic momenta. a)** and **b)** Fermi surfaces of LBCO at x=0.095 and x=0.165, respectively. **c)** and **d)** corresponding intensity contours at ω= -0.4 eV. Arrows indicate the wave-vector $\mathbf{k}_0$ of a peak in MDC at ω=−0.4 eV measured from the zone center along the zone diagonal. **e)** magnitude of $\mathbf{k}_0$ where "vertical dispersion" occurs for LBCO from this study (circles), BSCCO from ref. [11] (squares) and for CCOC from ref. [13] (triangle). Shaded region marks (π/4, π/4) wave-vector, suggested in [11] to be an "ordering" vector.

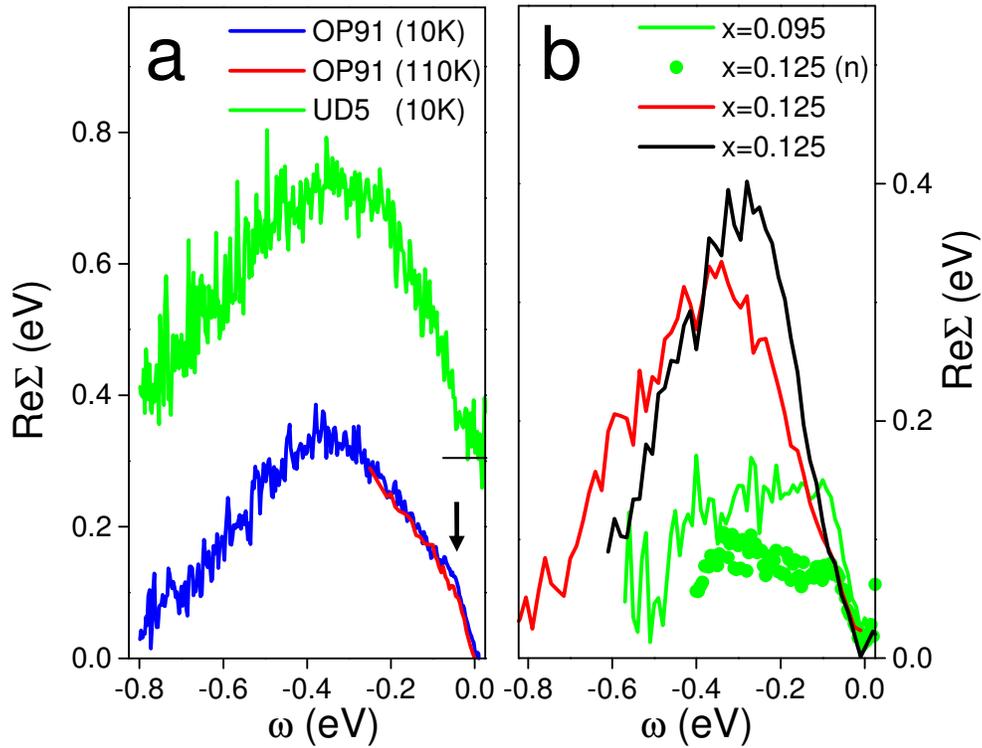

**Figure 3. ReΣ for samples/lines from fig. 1.** a) BCSSO and b) LBCO. Green line in a) is shifted up by 0.3 eV. Arrow indicates the low-energy region where changes occur between the normal and superconducting state. Green circles in b) represent the nodal ReΣ for x=0.125 LBCO (not shown in Fig. 1). Colors correspond to those used in Fig. 1.

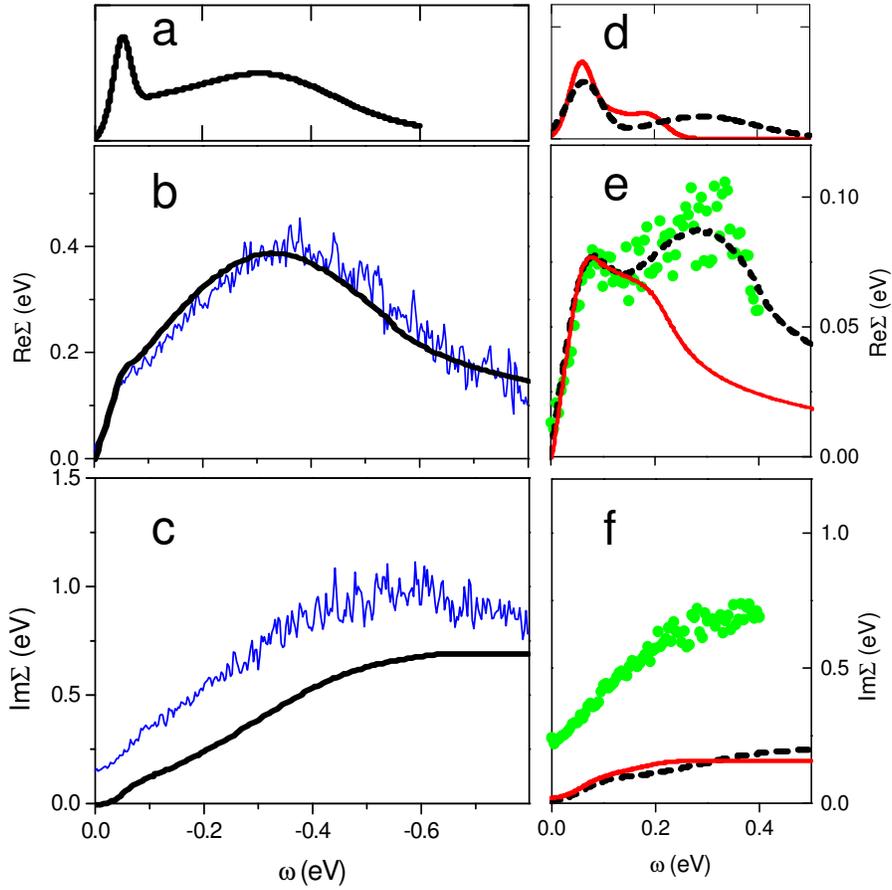

**Figure 4. A model for the excitation coupling spectrum. a)** A naive model for the excitation spectrum that produces self-energies similar to the measured ones in BSCCO **b)** ReΣ measured in Bi 2212 (thin lines) and model ReΣ obtained from the spectrum in a) (bold line). **c)** corresponding ImΣ. **d)** through **f)** Same for LBCO at x=1/8. Solid lines represent neutron scattering data from [16] and the self-energies derived from them. Dashed lines represent the excitation spectrum and derived self-energies that better model the high-energy region of measured self-energies (green circles).